\documentclass[twocolumn,letterpaper,amsmath,amssymb,floatfix,aps,superscriptaddress]{revtex4}

\usepackage{graphicx}
\usepackage{dcolumn}
\usepackage{bm}
\usepackage[usenames]{color}
\usepackage{hyperref}
\usepackage{ulem}
\usepackage{amsmath}
\usepackage{ulem}	% for put a line on text

\begin{document}

\title{Polymer translocation through nanopore assisted by an environment of active rods}

\author{Hamidreza Khalilian}
\email{khalilian@ipm.ir}
\affiliation{School of Nano Science, Institute for Research in Fundamental Sciences (IPM), 
19395-5531, Tehran, Iran.}

\author{Jalal Sarabadani}
\email{jalal@ipm.ir}
\affiliation{School of Nano Science, Institute for Research in Fundamental Sciences (IPM), 
19395-5531, Tehran, Iran.}

\author{Tapio Ala-Nissila}
\affiliation{Department of Applied Physics and QTF Center of Excellence, Aalto University, P.O. Box 11000, FI-00076 Aalto, Espoo, Finland.}
\affiliation{Interdisciplinary Centre for Mathematical Modelling and Department of Mathematical Sciences, 
Loughborough University, Loughborough, Leicestershire LE11 3TU, UK.}

\begin{abstract}
We use a combination of computer simulations and iso-flux tension propagation (IFTP) theory to investigate translocation dynamics of a flexible linear polymer through a nanopore into an environment composed of repulsive active rods in 2D. We demonstrate that the rod activity induces a crowding effect on the polymer, leading to a time-dependent net force that facilitates translocation into the active environment. Incorporating this force into the IFTP theory for pore-driven translocation allows us to characterise translocation dynamics in detail and derive a scaling form for the average translocation time as $\tilde{\tau} \sim \tilde{L}_{\textrm{r}}^{\nu} / \tilde{F}_{\textrm{SP}} $, where $\tilde{L}_{\textrm{r}}$ and $\tilde{F}_{\textrm{SP}}$ are the rod length and self-propelling force acting on the rods, respectively, and $\nu$ is the Flory exponent.
\end{abstract}

\maketitle

%%%%%%%%%%%%%%%%%%%%%%%%%%%%%%%%%%%%%%%%%%%%%%%%%%%%%%%%%%%%%%%%%%%%%%%%%%%%%%%%%%%%%%%%%%%%%%%%%%%%%%%%%%%%%%%%%%%%%%%%%%%%%%%%%%%%%%%%%%%%%%%
%%%%%%%%%%%%%%%%%%%%%%%%%%%%%%%%%%%%%%%%%%%%%%%%%%%%%%%%%%%%%%%%%%%%%%%%%%%%%%%%%%%%%%%%%%%%%%%%%%%%%%%%%%%%%%%%%%%%%%%%%%%%%%%%%%%%%%%%%%%%%%%

%
\begin{figure*}[t]\begin{center}
    \begin{minipage}{0.245\textwidth}
    \begin{center}
        \includegraphics[width=1.0\textwidth]{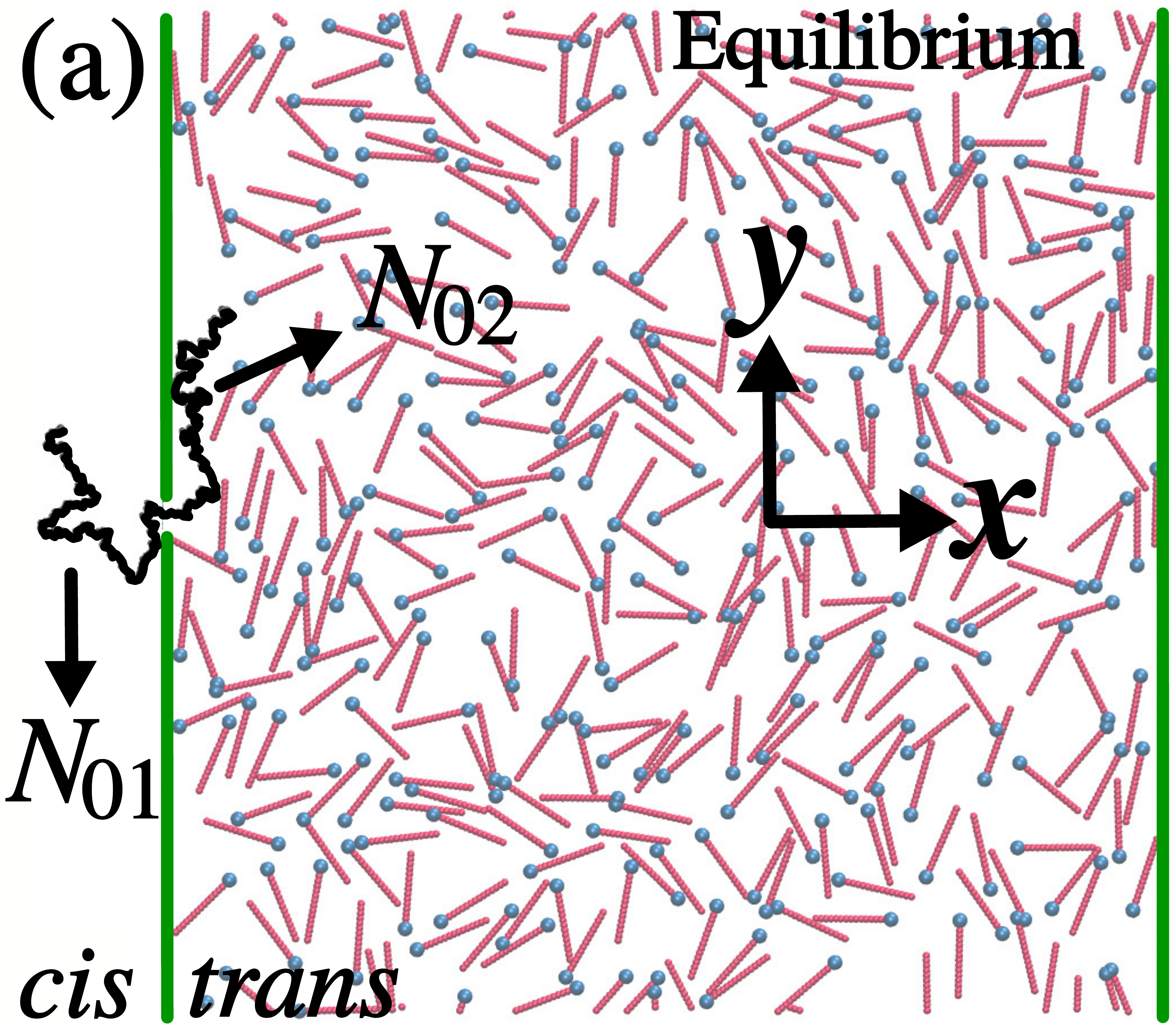}
    \end{center}\end{minipage} \hskip-0.0cm
        \begin{minipage}{0.245\textwidth}
    \begin{center}
        \includegraphics[width=1.0\textwidth]{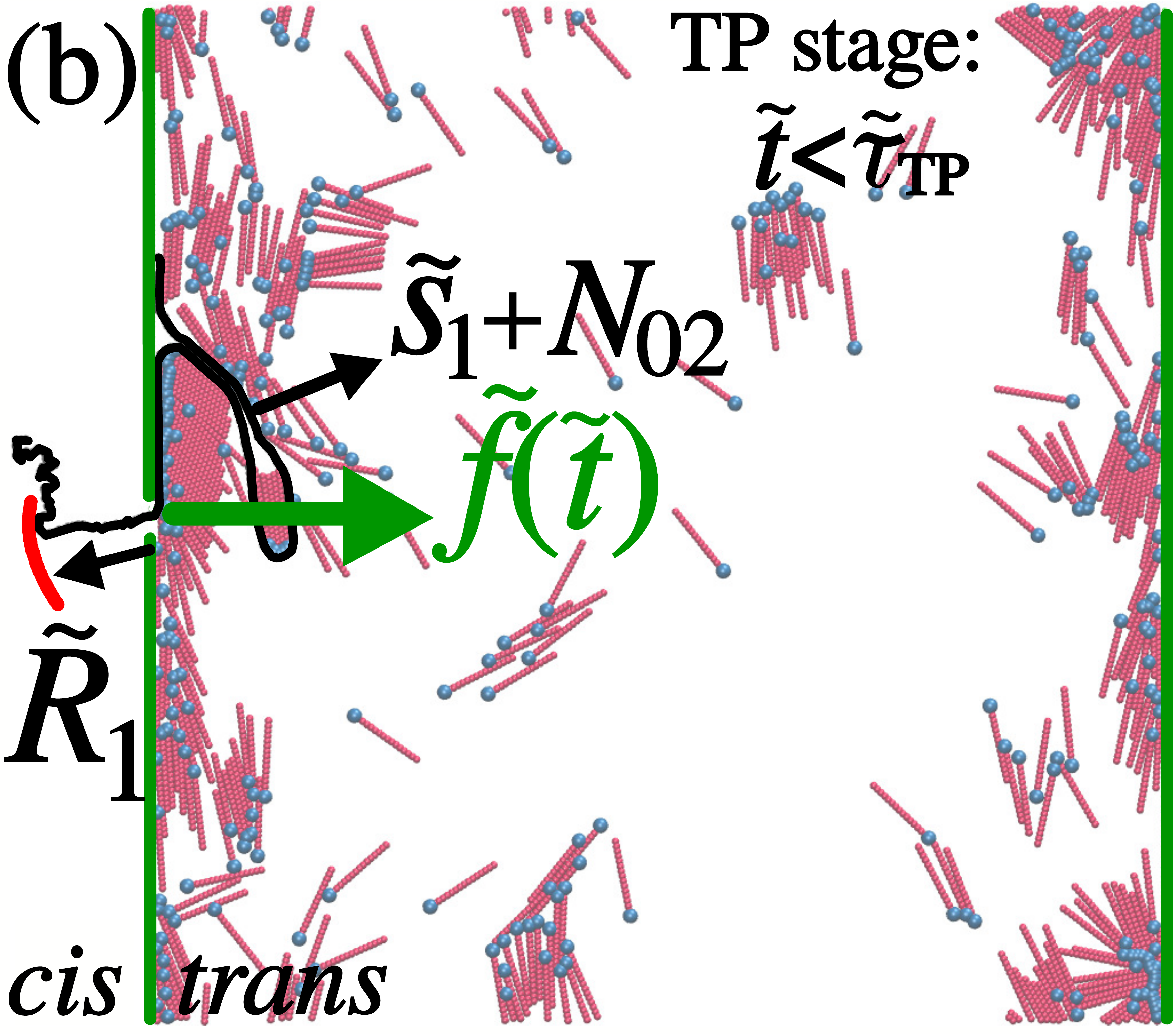}
    \end{center}\end{minipage} \hskip-0.0cm
        \begin{minipage}{0.245\textwidth}
    \begin{center}
        \includegraphics[width=1.0\textwidth]{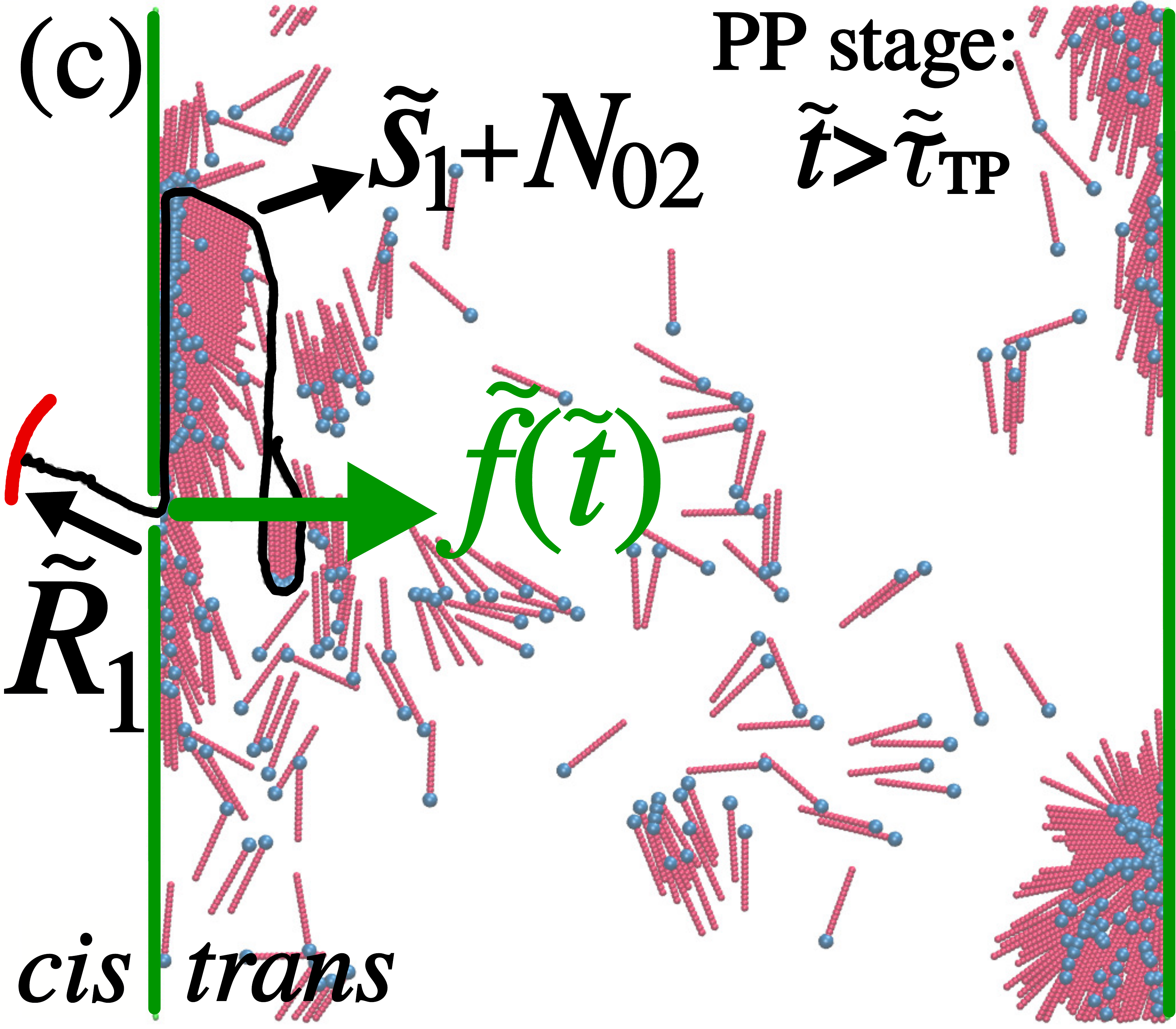}
    \end{center}\end{minipage} \hskip-0.0cm
    	\begin{minipage}{0.245\textwidth}
    \begin{center}
        \includegraphics[width=1.0\textwidth]{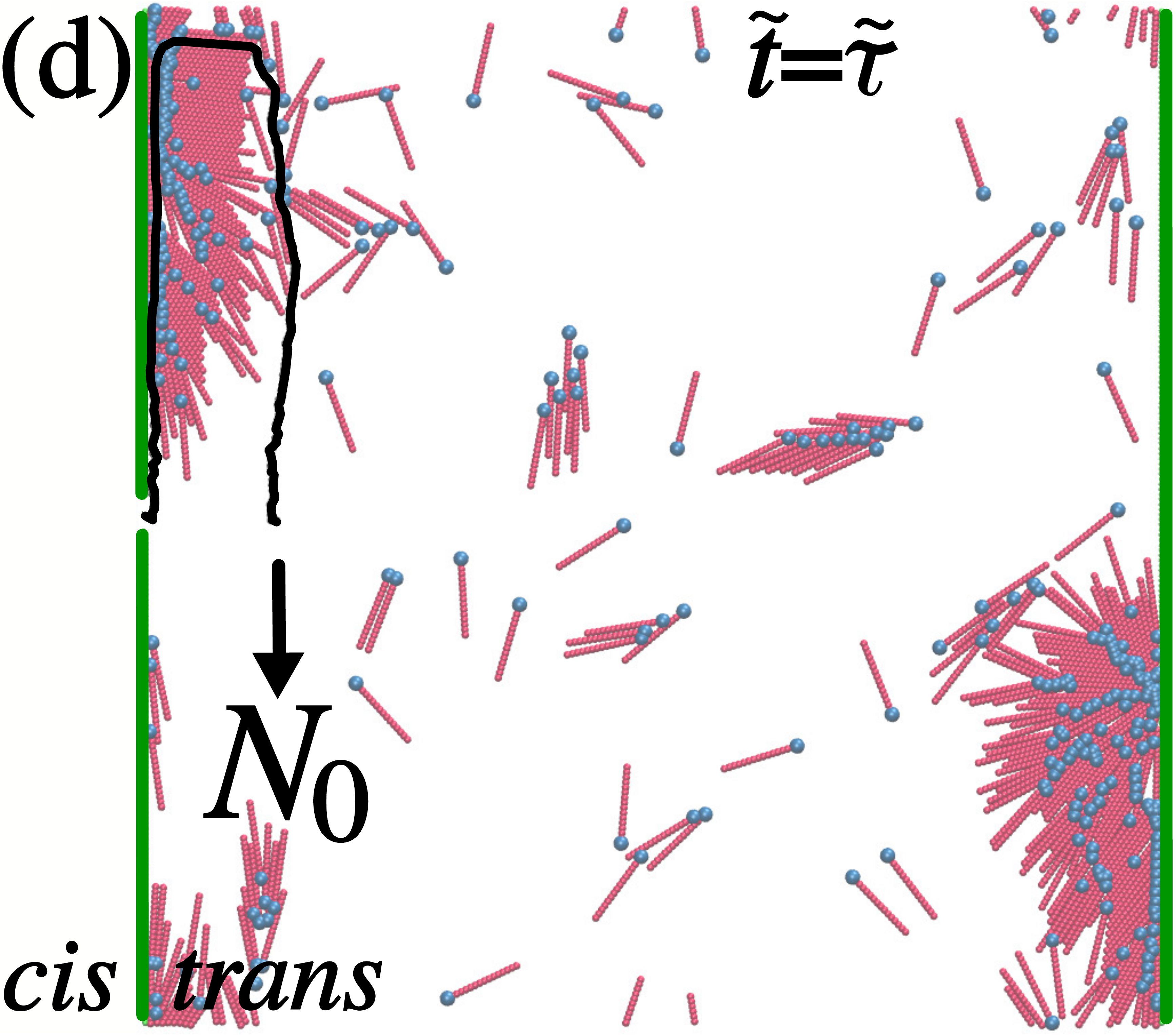}
    \end{center}\end{minipage}
\caption{(a) Configuration of the system after equilibration with passive rods with the chain fixed inside the pore. The compartment dimensions are $L_x = 2L_y = 400$ (equally partitioned by a membrane in the $y$ direction), $L_{\rm r} = 16$ (16 beads), and the number of rods $N_{\rm r} = 320$ which gives a number density of 0.128. Translocation is initiated from this state at $\tilde t=0$ by making the rods active with $F_{\textrm{SP}} = 32$ and releasing the chain. $N_{01}$ and $N_{02}$ are the initial contour lengths of the {\it cis} and {\it trans}-side subchains, respectively. (b) A snapshot of the system during the tension propagation (TP) stage, i.e. $\tilde{t} < \tilde{\tau}_{\textrm{TP}}$. The location of the tension front is denoted by $\tilde{R}_1$, and $\tilde{s}_1 + N_{02}$ is the translocation coordinate (the number of monomers on the {\it trans} side).  The SP force $F_{\textrm{SP}}$ acts on the blue head monomer of each rod and is directed parallel to its axis from tail to head. The effective force imposed by the rods on the chain $\tilde{f} (\tilde{t})$  originates from the interaction between the active rods and the {\it trans}-side subchain. (c) The same as panel (b) but for the post propagation stage, i.e. $\tilde{t} > \tilde{\tau}_{\textrm{TP}}$, where the tension has already reached the {\it cis}-side subchain end. (d) Final configuration of the system at the end of the translocation process.
}
\label{fig:schematic}
\end{center}
\end{figure*}
%

%%%%%%%%%%%%%%%%%%%%%%%%%%%%%%%%%%%%%%%%%%%%%%%%%%%%%%%%%%%%%%%%%%%%%%%%%%%%%%%%%%%%%%%%%%%%%%%%%%%%%%%%%%%%%%%%%%%%%%%%%%%%%%%%%%%%%%%%%%%%%%%%%%%%%%%%%%%%%%%%%%%%%%%%%%%%%%%%%%%%%%%%%%%%%%%%%%%%%%%%%%%%%%%%%%%%%%%%%%%%%%%%%%%%%%%%%%%%%%%%%%%%%%%%%%%%%%%%%%%%%%%%%%%%%%%%%%%%%%%%%%%%%%%%%%%%%%%%%%%%%%%%%%%%%%%%%%%%%%%%%%%%%%%

%{\it Introduction} -- 
Translocation dynamics of biopolymers through nanopores has been one of the most active research areas in soft matter during the last few decades \cite{Muthukumar_book,Tapio_review,MilchevJPCM2011,jalalJPCM2018}. The most relevant examples include DNA and mRNA translocation through nuclear pores, protein transportation across a membrane, and DNA injection by a virus. There are many applications from DNA sequencing to gene therapy and controlled drug delivery \cite{Alberts}, and forced translocation has been originally suggested as an inexpensive and fast method for DNA sequencing. Motivated by these applications many experimental as well as theoretical works \cite{mellerPRL2001,Smith_Nature_2001,Storm2003,BrantonPRL2003,StormNanoLett2005,Keyser_2006,%
Keyser_2009,Bulushev_2015,SungPRL1996,MuthukumarJCP1999,%
Kantor_PRE2004,GrosbergPRL2006,aksimentievNanolett2008,rowghanian2011,%
SakauePRE2007,SakauePRE2010,SakauePRE2012,TapioPRE2007,TapioEPJE2009,%
ikonen2012a,ikonen2012b,ikonen2013,ikonen2012c,jalalJCP2014,%
jalalJCP2015,JalalEPL2017,jalalSciRep2017,slaterPRE2010,slaterJCP2010,slaterPRE2009,hamidJCP2013,%
jalalPolymers2018,jalalPolymers2019,jalalJPCM2020_1,jalalJPCM2020_2,golestanianPRL2011} 
have been performed since the seminal works by Bezrukov {\it et al.} \cite{Bezrukov} and by Kasianowicz {\it et al.} \cite{KasiPNAS1996}. 
To date, most studies have focused on dynamics of polymer translocation facilitated by external driving in the pore, or by pulling the polymer from the head bead by optical tweezers, both of which are experimentally feasible \cite{SakauePRE2007,SakauePRE2010,SakauePRE2012,TapioPRE2007,TapioEPJE2009,%
ikonen2012a,ikonen2012b,ikonen2013,ikonen2012c,jalalJCP2014,%
jalalJCP2015,JalalEPL2017,jalalSciRep2017,slaterPRE2010,slaterJCP2010,slaterPRE2009}.

In biological systems, however, polymer translocation processes often occur in crowded environments \cite{Alberts}. 
Such environments may be composed of diffusive and randomly distributed spherical static obstacles \cite{gopinathan,chen,yu,samadi}, or chaperones that assist transport across membranes \cite{abdolvahab,emamyari, xu}. Crowded environments consisting of active particles (APs) have introduced a new out-of-equilibrium-dynamics field of research with rich physics \cite{vicsek,vicsek_rev}. Examples include synthetic motile objects from molecular scale to microns \cite{kay,weber}, microscopic living organisms \cite{berg,okubo}, and artificial swimmers from nano to millimeter scales \cite{dreyfus,tierno,leoni,purcell,najafi,avron,pooley,golestanian,deseigne1,%
deseigne2,blair,narayan,aranson,kudrolli1,snezhko,kudrolli2,daniels}. The presence of APs has strong influence on polymer chains in equilibrium and can lead to significant conformational and dynamical changes (collapse or swelling)  
depending on the system parameters  \cite{kaiser, harder, Chelakkot, jiang2014_1, jiang2014_2, liu2019, liu2020}.

An interesting open question pertains to the influence of APs on polymer translocation dynamics. Pore-driven polymer translocation in the presence of spherical APs has been considered using computer simulation methods \cite{pu} and it was found that with high activity, there's a crowding effect in two dimensions (2D) that leads to a speed-up of translocation. An interesting and unanswered question remains concerning the effect of the APs on dynamics of unforced or non-driven 
translocation. In many cases active objects such as bacteria are not spherical, but assume rodlike shapes. Combination of anisotropy in the shape of the APs and the presence of a self-propelling (SP) force that makes the objects active leads into interesting collective dynamics. In the case of rodlike APs, there is orientational alignment of the ARs along the walls in a 2D confining channel \cite{lowen},  and also in their collective motion \cite{peruani}. The presence of a translocating polymer in such an environment thus warrants closer examination.

To this end, in this Letter we perform extensive computer simulations of unforced polymer translocation dynamics in the presence of active rods (ARs) on the {\it trans} side of the pore. We demonstrate that when the chain is initially placed in such a way that a part of the chain is in the {\it trans} compartment, the presence of ARs induces a net force from the {\it cis} to the {\it trans} side that overcomes entropic losses and facilitates translocation. The presence of a time-dependent driving force allows us to use the iso-flux tension propagation (IFTP) theory which is benchmarked against the simulation data. As our main theoretical result, we find that the mean translocation time $\tilde{\tau}$ scales with the rod length $\tilde{L}_{\textrm{r}}$ and SP force $\tilde{F}_{\textrm{SP}}$ as $\tilde{\tau} \sim \tilde{L}_{\textrm{r}}^{\nu} / \tilde{F}_{\textrm{SP}} $, where $\nu$ is the equilibrium Flory exponent.

%%%%%%%%%%%%%%%%%%%%%%%%%%%%%%%%%%%%%%%%%%%%%%%%%%%%%%%%%%%%%%%%%%%%%%%%%%%%%%%%%%%%%%%%%%%%%%%%%%%%%%%%%%%%%%%%%%%%%%%%%%%%%%%%%%%%%%%%%%%%%%%%%%%%%%%%%%%%%%%%%%%%%%%%%%%%%%%%%%%%%%%%%%%%%%%%%%%%%%%%%%%%%%%%%%%%%%%%%%%%%%%%%%%%%%%%%%%%%%%%%%%%%%%%%%%%%%%%%%%%%%%%%%%%%%%%%%%%%%%%%%%%%%%%%%%%%%%%%%%%%%%%%%%%%%%%%%%%%%%%%%%%%%%%%%%%%%%%%%%%%%%%%%%%%%%%%%%%%%%%%%%%%%%%%%%%%%%%%%%%%%%%%%%%%%%%%%%%%%%%%%%%%%%%%%%%%%%%%%%%%%%%%%%%%%%%%%%%

Our system comprises a flexible, self-avoiding polymer chain of length $N_0$ modeled by the bead-spring model \cite{grest} with beads having a pairwise shifted repulsive Lennard-Jones (SRLJ) interaction $U_{\textrm{LJ}}(r) = 4 \varepsilon \big[ (\frac{\sigma}{r})^{12}-(\frac{\sigma}{r})^{6}\big] + \varepsilon $ if $r\leq2^{1/6}$, and zero otherwise. Here $\sigma$ is the LJ radius, $\varepsilon$ is the potential well depth and $r$ is the distance between two monomers. In addition, the consecutive monomers are connected by the finitely extensible nonlinear elastic (FENE) interaction $U_{\textrm{FENE}}(r) = -\frac{1}{2}k R^{2}_{0}\ln[1-r^2 / R_0^2]$, where $k$ and $R_0$ are the string constant and the maximum allowed distance between the consecutive monomers, respectively. The polymer is put into a container of size $L_x = 2L_y = 400$ in units of $\sigma$, and there is a membrane in the middle (see Fig.~\ref{fig:schematic}) with a nanopore of radius $1.5 \sigma$ allowing only one bead in the pore at a time. The container has walls in the $x$ direction and periodic boundary conditions in the $y$ direction, and its walls and the membrane interact with the chain with the same SRLJ potential. In the {\it trans} side there are $N_{\textrm{r}}$ rigid rods comprising SRLJ beads of radius $\sigma$ whose length is $L_{\textrm{r}} \sigma$. To model self-propulsion, a SP force with magnitude of $F_{\textrm{SP}}$ is added to the head bead of each rod along its main axis and from its tail to its head. 

For the simulations we employ Langevin dynamics (LD), where for the position of the $i^{\rm th}$ monomer of the polymer $ {M} \ddot{\vec{r}}_{i} = -\eta \dot{\vec{r}}_{i} - \vec{\nabla} U_{\textrm{m}i} + \vec{\xi}_{i}(t)$. Here $\eta$ is the friction coefficient, $U_{\textrm{m}i}$ is the sum of all interactions, and $\vec{\xi}_i$ is white noise with $\langle \vec{\xi}_{i} (t) \rangle  =0$ and $\langle \vec{\xi}_{i} (t) \vec{\xi}_{j} (t') \rangle  = 2 \eta k_{\textrm{B}} T \delta_{ij} \delta (t - t')$, with $k_{\textrm{B}}$ the Boltzmann constant, $T$ the temperature, and $\delta_{ij}$ and $\delta (t -t')$ are the Kronecker and Dirac delta functions, respectively.
For the $i^{\rm th}$ bead of each rod we add the SP force as ${M} \ddot{\vec{r}}_{i} = -\eta \dot{\vec{r}}_{i} + F_{\textrm{SP}} \delta_{ih} \hat{\textrm{e}} - \vec{\nabla} U_{\textrm{r}i} + \vec{\xi}_{i}(t)$, where $h$ picks the head bead, $\hat{\textrm{e}}$ is unit vector parallel to the vector connecting the tail to the head, and $U_{\textrm{r}i}$ is the sum of all interactions on the $i^{\rm th}$ bead. We use $M, \sigma$ and $\varepsilon$ as the units for mass, length and energy, respectively, where $M=1$ is mass of each monomer in the polymer and the rods, and $\varepsilon=1$. The temperature is kept at $k_{\textrm{B}}T=1.2$, the solvent friction coefficient is $\eta=0.7$, and $\tau_{0}=\sqrt{M\sigma^{2}/\varepsilon}$ is the simulation time unit. In our simulations, the integration time step is $\textrm{d} t=0.001 \tau_{0}$. Finally, the spring constant is set to $k=30$ and $R_{0}=1.5$. The simulations are performed using LAMMPS \cite{lammps} package. 

Before the translocation process, the polymer is fixed in the pore such that there are $N_{01}$ and $N_{02}$ beads in the {\it cis} and {\it trans} compartments, respectively, where $N_0=1+N_{01}+N_{02}$. The polymer-rod system is equilibrated for $t_{\textrm{eq}}= 5 \times 10^{4}\tau_{0}$ without the SP force, and at the beginning of translocation the polymer is released and $F_{\textrm{SP}}$ turned on for all rods simultaneously. The density of the ARs here has been chosen low enough such that the equilibrated system has a uniformly random orientational distribution. We use dimensionless quantities throughout and tilde denotes units within the IFTP theory \cite{dimensionless}.

%%%%%%%%%%%%%%%%%%%%%%%%%%%%%%%%%%%%%%%%%%%%%%%%%%%%%%%%%%%%%%%%%%%%%%%%%%%%%%%%%%%%%%%%%%%%%%%%%%%%%%%%%%%%%%%%%%%%%%%%%%%%%%%%%%%%%%%%%%%%%%%%%%%%%%%%%%%%%%%%%%%%%%%%%%%%%%%%%%%%%%%%%%%%%%%%%%%%%%%%%%%%%%%%%%%%%%%%%%%%%%%%%%%%%%%%%%%%%%%%%%%%%%%%%%%%%%%%%%%%%%%%%%%%%%%%%%%%%%%%%%%%%%%%%%%%%%%%%%%%%%%%%%%%%%%%%%%%%%%%%%%%%%%%%%%%%%%%%%%%%%%%%%%%%%%%%%%%%%%%%%%%%%%%%%%%%%%%%%%%%%%%%%%%%%%%%%%%%%%%%%%%%%%%%%%%%%%%%%%%%%%%%%%%%%%%%%%%

Figure~\ref{fig:schematic}(a) presents a typical snapshot of the system after equilibration at $\tilde{t} = 0$.  We find that the activity of the rods induces an effective force $\tilde{f} (\tilde{t})$ on the monomer in the pore, directed from {\it cis} to {\it trans} such that it facilitates translocation as depicted in panel (b). This causes tension propagation along the backbone of the {\it cis}-side subchain. The distance between the tension front in the {\it cis} side from the pore is denoted by $\tilde{R}_1$. This stage is called the tension propagation (TP) stage ($\tilde{t} < \tilde{t}_{\textrm{TP}}$). The number of the monomers on the {\it trans} side $\tilde{s}_1 + N_{02}$ defines a translocation coordinate, and is equal to $N_{02}$ at time zero. Eventually the tension reaches the end of the {\it cis}-side subchain and the post propagation (PP) stage starts ($\tilde{t} > \tilde{t}_{\textrm{TP}}$), which is presented in panel (c). Finally, panel (d) shows last snapshot of the system at the end of the translocation process which defines the translocation time $\tilde \tau$.
\begin{figure}[t]\begin{center}
    \begin{minipage}{0.38\columnwidth}
    \begin{center}
        \includegraphics[width=1.0\textwidth]{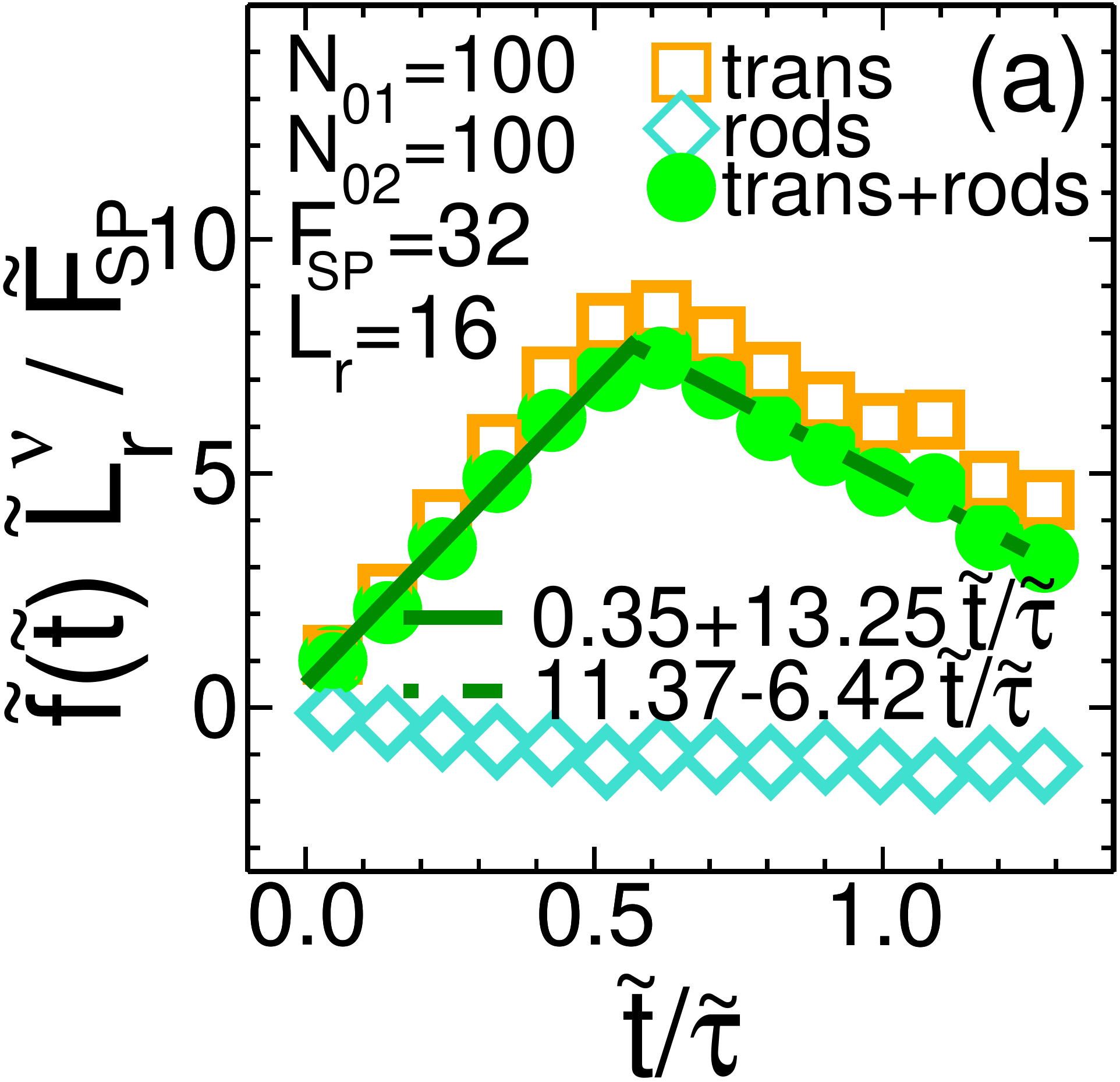}
    \end{center}\end{minipage} \hskip-0.1cm
    \begin{minipage}{0.3\columnwidth}
    \begin{center}
        \includegraphics[width=1.0\textwidth]{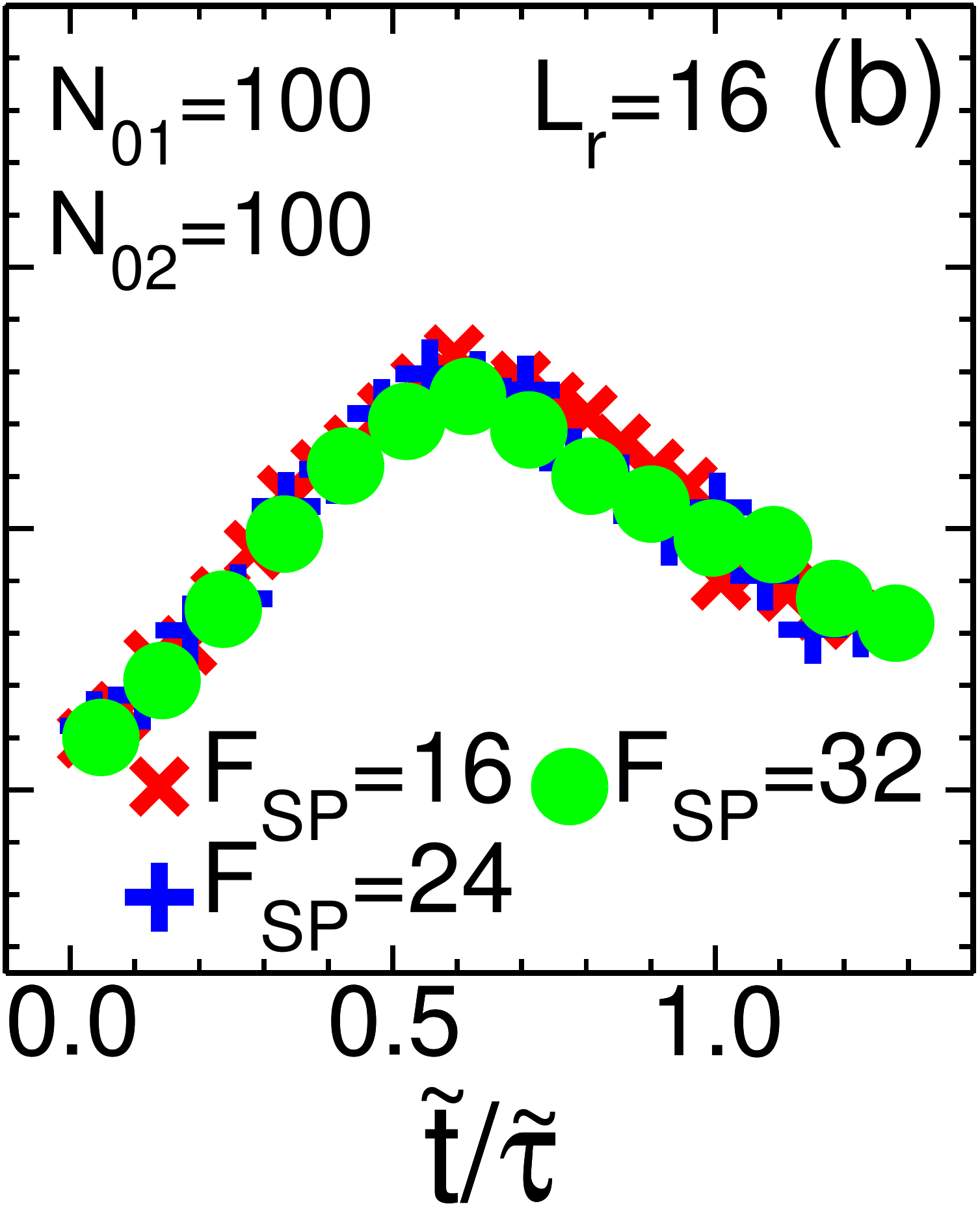}
    \end{center}\end{minipage} \hskip-0.1cm
    \begin{minipage}{0.3\columnwidth}
    \begin{center}
        \includegraphics[width=1.0\textwidth]{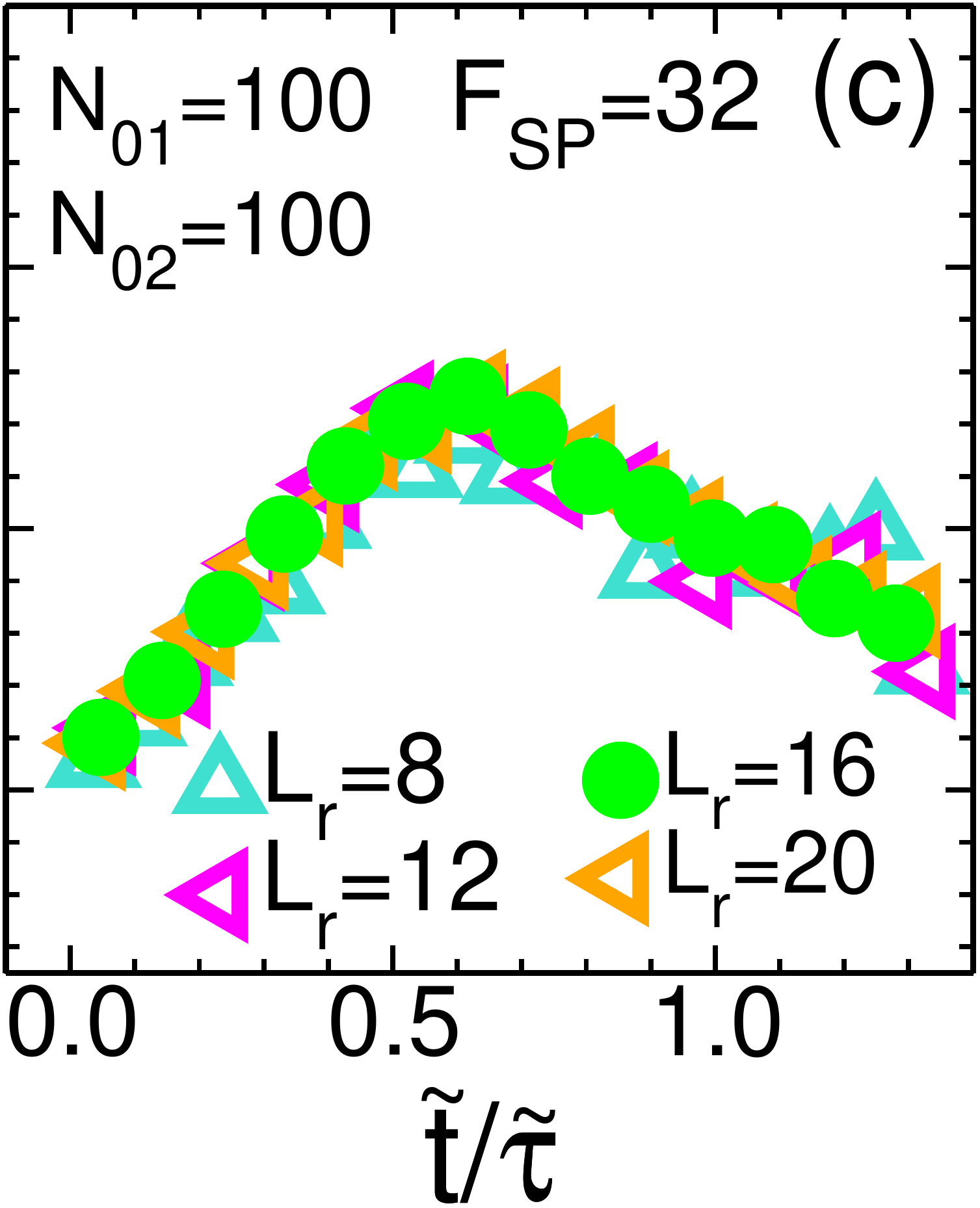}
    \end{center}\end{minipage}
\caption{(a) Normalized effective force in the $x$ direction from the LD simulations $\tilde{f} (\tilde{t})\tilde{L}_{\textrm{r}}^{\nu} / \tilde{F}_{\textrm{SP}}$ experienced by the monomer inside the pore as a function of the normalized time $\tilde{t} / {\tilde \tau}$, where $\nu$ is the Flory exponent and $\tilde \tau$ is the translocation time (as obtained from LD) for $L_{\textrm{r}} = 16$, $N_{\textrm{r}} = 320$ and $F_{\textrm{SP}} = 32$. Open orange squares and open turquoise diamonds are data for the contributions of the tension due to the {\it trans}-side subchain and the interactions of ARs with the monomer inside the pore, respectively. The total force which is the sum of the tension due to the {\it trans}-side subchain and the AR interactions with the monomer inside the pore is given by green circles. The solid and the dashed green lines are the fitting curves for the total effective force in the TP and PP stages, respectively. $N_{01}=100$ and $N_{02}=100$ are the initial contour lengths of the {\it cis} and {\it trans}-side subchains, respectively. (b) The same as panel (a) but for fixed $L_{\textrm{r}}  = 16$ and for different values of $F_{\textrm{SP}} = 16, 24$ and 32. (c) The same as panel (b) but for fixed $F_{\textrm{SP}} =32$ and different values of $L_{\textrm{r}}  = 8, 12, 16$ and 20 corresponding to the number of rods $N_{\textrm{r}} = 640, 426, 320$ and 256, respectively. 
}
\label{fig:forceMD_BLength}
\end{center}
\end{figure}

Over the last few years a consistent and quantitatively accurate theory of driven polymer translocation has been developed based on iso-flux tension propagation (IFTP). Due to the effective force  $\tilde{f} (\tilde{t})$ induced by the SP forces of the rods, we can generalize the IFTP theory to the present case, too. To obtain the time evolution of the translocation coordinates $\tilde{s}_{1}$, the iso-flux (IF) approximation for the monomer flux $\tilde{\phi}_1 (\tilde{t}) = \textrm{d} \tilde{s}_1 / \textrm{d} \tilde{t} $ means that it is constant in space but evolves in time within the mobile subchain on the {\it cis} side.
The force at distance $\tilde{x}$ from the pore on the {\it cis} side $\tilde{f} (\tilde{x} , \tilde{t})$ is obtained by integration of the local force balance relation $\textrm{d} \tilde{f} (\tilde{x}') = - \tilde{\phi}_1 \textrm{d} x' $ from the pore entrance to the distance $\tilde{x}$ as  $\tilde{f} (\tilde{x} , \tilde{t}) = \tilde{f}_0 - \tilde{x} \tilde{\phi}_1 (\tilde{t})$ (here the force at the entrance of the pore on the {\it cis} side is $\tilde{f}_0 = \tilde{f} (\tilde{t}) - \tilde{\eta}_{\textrm{p}} \tilde{\phi}_1 (\tilde{t}) $). Then as the tension force vanishes at $\tilde{R}_1$ i.e. $\tilde{f} (\tilde{R}_1) = 0$, the equation of motion for the translocation coordinates is cast into \cite{jalalSciRep2017,JalalEPL2017}
\begin{equation}
\tilde{\Gamma}_1 (\tilde{t}) \frac{ \textrm{d} \tilde{s}_1 }{ \textrm{d} \tilde{t} } = 
\tilde{f} (\tilde{t}) ,
\label{BD_force}
\end{equation}
where $\tilde{\Gamma}_1 (\tilde{t}) = \tilde{R_1}(\tilde{t}) + \tilde{\eta}_{\textrm{p}}$, with $ \tilde{\eta}_{\textrm{p}}$ and $\tilde{R}_1 (\tilde{t})$ as the pore friction and the tension front distance from the nanopore, respectively.

In Fig.~\ref{fig:forceMD_BLength}(a) we plot the total normalized effective force $\tilde{f} (\tilde{t}) \tilde{L}_{\textrm r}^{\nu} /\tilde{F}_{\textrm{SP}} $ (filled green circles) in the horizontal direction from {\it cis} to {\it trans} as a function of the normalized time $\tilde{t} / {\tilde \tau}$ for  $N_{01} = N_{02} = 100$, $F_{\textrm{SP}} = 32$ and $L_{\textrm{r}} = 16$. It is the sum of the normalized tension force due to the {\it trans}-side subchain (open orange squares) and the normalized force due to the interactions of the ARs with the monomer inside the pore (open turquoise diamonds). The force first grows almost linearly and then decreases. Its largest magnitude in LJ units is of the same order as that of $F_{\textrm{SP}}$. In panels (b) and (c) we show the corresponding data for varying either $F_{\rm SP}$ or $L_{\rm r}$. The remarkable finding here is that all the data collapse on two master curves, namely $\tilde{f}_{\textrm{TP}} (\tilde{t}) \tilde{L}_{\textrm r}^{\nu} /\tilde{F}_{\textrm{SP}} = 0.35 + 13.25 \tilde{t} / \tilde{\tau} $ (solid green line in panel (a)) and $ \tilde{f}_{\textrm{PP}} (\tilde{t}) \tilde{L}_{\textrm r}^{\nu} /\tilde{F}_{\textrm{SP}} = 11.37 - 6.42 \tilde{t} / \tilde{\tau}$ (dashed-dotted green line in panel (a)) in the TP and PP stages, respectively, and they intersect at $\tilde t / \tilde \tau = 0.56$. We have independently verified from the bond lengths that the maximum of the force curve exactly corresponds to the TP time where the tension front reaches the end of the {\it cis}-side subchain.

Solving Eq.~(\ref{BD_force}) gives the translocation coordinate $\tilde{s}_1$ provided that the time evolution of the location of the tension front $\tilde{R}_1$ is known. To obtain the equation of motion for $\tilde{R}_1$ in the TP stage, the corresponding closure relation $\tilde{R}_1 = A_{\nu} N_1^{\nu}$ must be used. Here $\nu = 3/4$ is the Flory exponent in 2D, and $N_1=\tilde{l}_1 + \tilde{s}_1$ is the number of monomers in the {\it cis}-side subchain that have been influenced by the tension force, and $\tilde{l}_1$ is the number of monomers in the mobile domain on the {\it cis} side.
Assuming that the mobile part of the {\it cis}-side subchain is fully straightened corresponding to the strong stretching (SS) regime of polymer translocation dynamics, we can write $\tilde{l}_1 = \tilde{R}_1$ \cite{rowghanian2011,jalalJCP2014}.  
Together with the definition of the monomer flux $\tilde{\phi}_1 = \textrm{d} \tilde{s}_1 / \textrm{d} \tilde{t}$, and differentiating both sides of $\tilde{R}_1 = A_{\nu} (\tilde{R}_1 + \tilde{s}_1)^{\nu}$ in time, the equation of motion for $\tilde{R}_1$ in the TP stage is
\begin{equation}
\dot{\tilde{R}}_1 (\tilde{t}) =  \frac{ \nu A_{\nu}^{1/\nu} \tilde{R}_1 (\tilde{t}) ^{(\nu-1)/\nu} \tilde{\phi}_1 (\tilde{t}) }{ 1- \nu A_{\nu}^{1/\nu} \tilde{R}_1 (\tilde{t}) ^{(\nu-1)/\nu}  } .
\label{R_TP}
\end{equation}
In the PP stage as the tension has already reached the {\it cis}-side subchain. Differentiating the closure $N_1 = \tilde{l}_1 + \tilde{s}_1 = N_{01}$ gives the time evolution of $\tilde{R}_1$ as
\begin{equation}
\dot{\tilde{R}}_1 (\tilde{t}) =  - \tilde{\phi}_1.
\label{R_PP}
\end{equation}
To have the full solution of the IFTP theory in the TP stage both Eqs.~(\ref{BD_force}) and (\ref{R_TP}) must be self-consistently solved, while in the PP stage one has to solve Eqs.~(\ref{BD_force}) and (\ref{R_PP}).

%%%%%%%%%%%%%%%%%%%%%%%%%%%%%%%%%%%%%%%%%%%%%%%%%%%%%%%%%%%%%%%%%%%%%%%%%%%%%%%%%%%%%%%%%%%%%%%%%%%%%%%%%%%%%%%%%%%%%%%%%%%%%%%%%%%%%%%%%%%%%%%%%%%%%%%%%%%%%%%%%%%%%%%%%%%%%%%%%%%%%%%%%%%%%%%%%%%%%%%%%%%%%%%%%%%%%%%%%%%%%%%%%%%%%%%%%%%%%%%%%%%%%%%%%%%%%%%%%%%%%%%%%%%%%%%%%%%%%%%%%%%%%%%%%%%%%%%%%%%%%%%%%%%%%%%%%%%%%%%%%%%%%%%%%%%%%%%%%%%%%%%%%%%%%%%%%%%%%%%%%%%%%%%%%%%%%%%%%%%%%%%%%%%%%%%%%%%%%%%%%%%%%%%%%%%%%%%%%%%%%%%%%%%%%%%%%%%%

\begin{figure}[t]\begin{center}
        \begin{minipage}{0.64\columnwidth}
    \begin{center}
        \includegraphics[width=1.0\columnwidth]{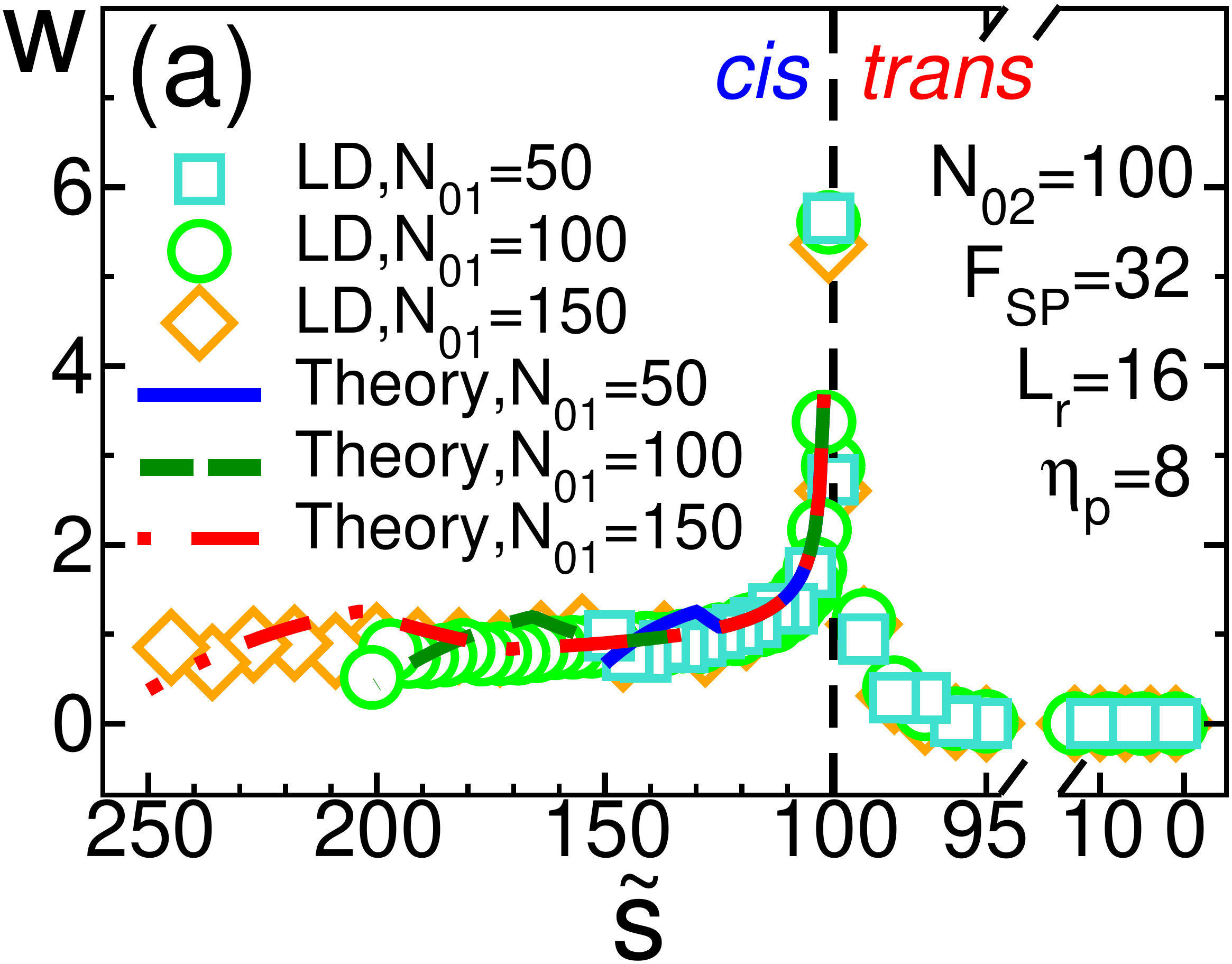}
    \end{center}\end{minipage}  \hspace{-0.2cm}
    \begin{minipage}{0.34\columnwidth}
    \begin{center}
        \includegraphics[width=1.0\columnwidth]{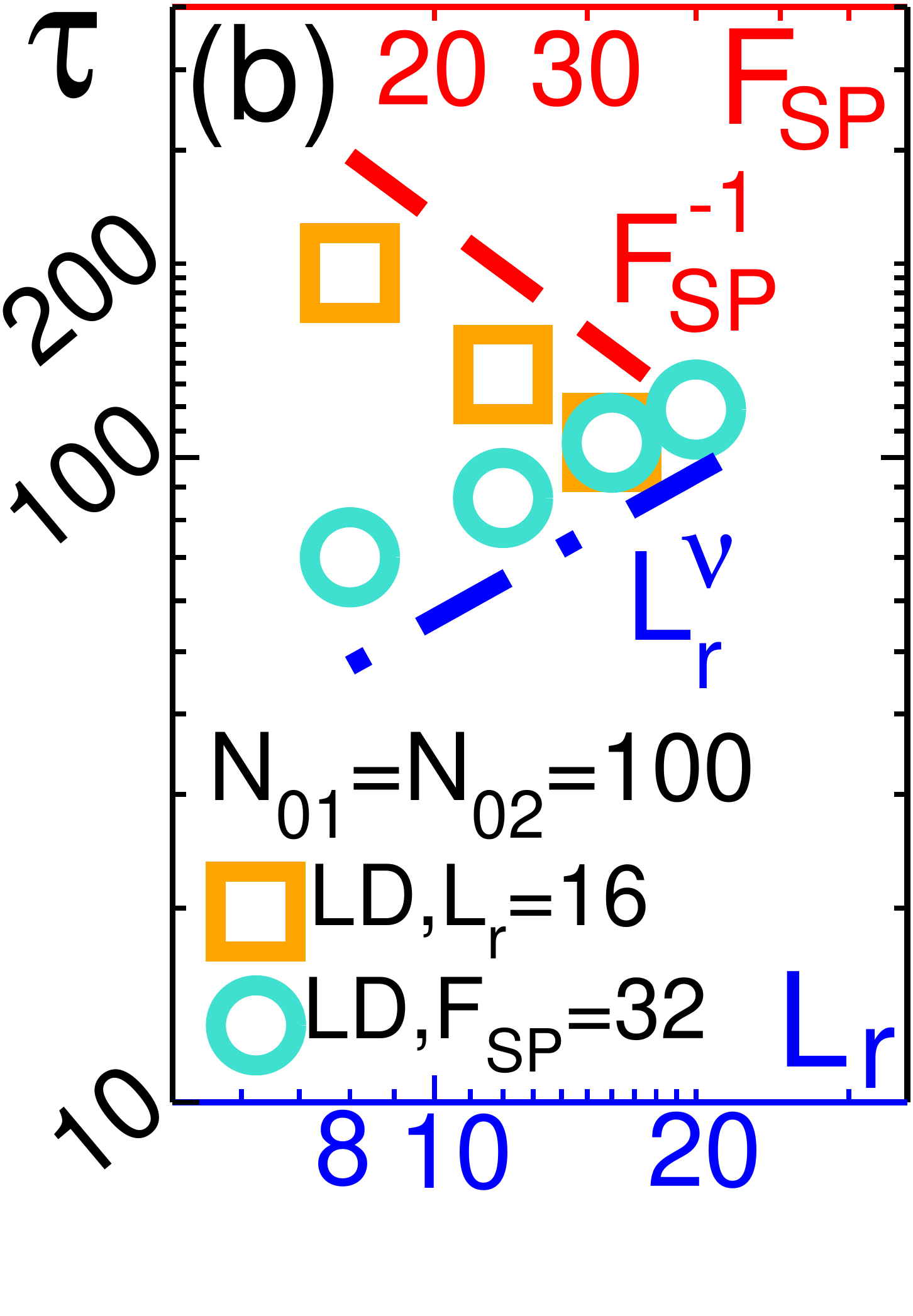}
    \end{center}\end{minipage} 
\caption{(a) Waiting time $w$ as a function of the total translocation coordinate $\tilde{s}$ for fixed values of the initial {\it trans}-side contour length $N_{02}=100$, SP force $F_{\textrm{SP}}=32$, rod length $L_{\textrm{r}} = 16$, and pore friction in the IFTP theory $\eta_{\textrm{p}}=8$, and for different values of the initial {\it cis}-side contour lengths $N_{01}=50$ (open turquoise squares), $N_{01}=100$ (open green circles) and $N_{01}=150$ (open orange diamonds). The solid blue, dashed green and dashed-dotted red lines are the IFTP results for $N_{01}=50, 100$ and 150, respectively. (b) Translocation time $\tau$ (from LD simulations) plotted as a function of rod length $L_{\textrm{r}}$ for fixed $F_{\textrm{SP}}=32$ (open turquoise circles from the bottom horizontal blue axis).  Open orange squares are the LD data for the translocation time as a function of $F_{\textrm{SP}}$ (from the top horizontal red axis) for fixed  $L_{\textrm{r}} =16$. Dashed red and dashed-dotted blue lines are guides to the eye.
}
\label{waiting}
\end{center}
\end{figure}

To validate the IFTP theory it is useful to investigate the waiting time (WT) distribution $w$, which is the time that each bead spends in the pore during the translocation process. In Fig.~\ref{waiting}(a) we plot $w(\tilde s)$ as a function of the total  translocation coordinate $\tilde{s}= \tilde{s}_1 + \tilde{s}_2$ ($\tilde{s}_1$ and $\tilde{s}_2$ correspond to the {\it cis} and {\it trans}-side subchains, respectively) for $N_{02}=100$, $F_{\textrm{SP}}=32$,  $L_{\textrm{r}} = 16$, and pore friction coefficient in the IFTP theory $\tilde \eta_{\textrm{p}}=8$ (which can be obtained by comparing WT from IFTP with LD simulations), and for different values of the initial {\it cis} side contour lengths $N_{01}=50$ (open turquoise squares), $N_{01}=100$ (open green circles) and $N_{01}=150$ (open orange diamonds). The solid blue, dashed green and dashed-dotted red lines present the IFTP results for $N_{01}=50, 100$ and 150, respectively. Regions with $ 0 < \tilde{s} \leq N_{02}$ and $N_{02} < \tilde{s} \leq N_0$ identify the monomers initially in the {\it trans} and the {\it cis}-side subchains, respectively. Here, $N_{02}=100$ has been fixed in order to have the same initial configuration for the {\it trans} side sub-system. This allows us to investigate the effect of the initial contour length only of the {\it cis}-side subchain on the translocation process. We find good agreement between the LD simulation results and the IFTP theory. The simulation data show that the {\it trans}-side subchain ($ 0 < \tilde{s} \leq N_{02}$) contributes to WT (see Fig.~\ref{fig:forceMD_BLength}) due to the small magnitude of $\tilde{f} (\tilde{t})$ at the beginning of translocation process because a short section of the {\it trans}-side subchain is 
temporarily retracted to the {\it cis} side. The IFTP theory thus slightly underestimates WT as it considers dynamics of the {\it cis}-side subchain only.

%

%%%%%%%%%%%%%%%%%%%%%%%%%%%%%%%%%%%%%%%%%%%%%%%%%%%%%%%%%%%%%%%%%%%%%%%%%%%%%%%%%%%%%%%%%%%%%%%%%%%%%%%%%%%%%%%%%%%%%%%%%%%%%%%%%%%%%%%%%%%%%%%%%%%%%%%%%%%%%%%%%%%%%%%%%%%%%%%%%%%%%%%%%%%%%%%%%%%%%%%%%%%%%%%%%%%%%%%%%%%%%%%%%%%%%%%%%%%%%%%%%%%%%%%%%%%%%%%%%%%%%%%%%%%%%%%%%%%%%%%%%%%%%%%%%%%%%%%%%%%%%%%%%%%%%%%%%%%%%%%%%%%%%%%%%%%%%%%%%%%%%%%%%%%%%%%%%%%%%%%%%%%%%%%%%%%%%%%%%%%%%%%%%%%%%%%%%%%%%%%%%%%%%%%%%%%%%%%%%%%%%%%%%%%%%%%%%%%%

Next we consider the average translocation time $\tilde{\tau}$ for the polymer needs to pass through the nanopore. Its scaling form can be written as $\tilde{\tau} \propto \tilde{F}_{\textrm{SP}}^{\beta} \tilde{L}_{\textrm{r}}^{\gamma} N_{01}^{\alpha} $, where $\alpha$, $\beta$ and $\gamma$ are the translocation, SP force and rod length exponents, respectively. Combining mass conservation in the TP and PP stages, i.e. $N_1= \tilde{s}_1 + \tilde{l}_1$ and $N_1= \tilde{s}_1+ \tilde{l}_1 = N_{01}$, respectively, with Eq.~(\ref{BD_force}), the TP time is obtained by integration of $N_1$ from zero to $N_{01}$, while the PP time is calculated by integration of $\tilde{R}_1$ from $\tilde{R}_1 (N_{01})$ to zero \cite{jalalJCP2014}. The sum of the TP and PP times leads to $\int_{0}^{\tilde{t}_{\textrm{TP}}} \!\!\! \tilde{f}_{\textrm{TP}} (\tilde{t}) \textrm{d} \tilde{t}
+ \int_{\tilde{t}_{\textrm{TP}}}^{\tilde{\tau}} \! \tilde{f}_{\textrm{PP}} (\tilde{t}) \textrm{d} \tilde{t} = \!\! \int_{0}^{N_{01}} \!\! \textrm{d} N_1 [ \tilde{R}_1 (N_1) + \tilde{\eta}_{\textrm{p}} ]$, where the effective forces in the TP and PP stages of the above relation are obtained from fitting to the simulation data in Fig.~\ref{fig:forceMD_BLength}(a) as $\tilde{f}_{\textrm{TP}} (\tilde{t})\tilde{L}_{\textrm{r}}^{\nu} / \tilde{F}_{\textrm{SP}} = a +b \tilde{t} / \tilde{\tau}$ (solid green line) and $\tilde{f}_{\textrm{PP}} (\tilde{t})\tilde{L}_{\textrm{r}}^{\nu} / \tilde{F}_{\textrm{SP}} = c -d \tilde{t} / \tilde{\tau}$ (dashed-dotted green line), respectively, with $a=0.35$, $b=13.25$, $c=11.37$ and $d=6.42$. The TP time $\tilde{f}_{\textrm{TP}} (\tilde{t} = \tilde{t}_{\textrm{TP}}) = \tilde{f}_{\textrm{PP}} (\tilde{t} = \tilde{t}_{\textrm{TP}}) $ is then
given by $\tilde{t}_{\textrm{TP}} = Q \tilde{\tau}$, where $Q=  (c-a) / (b+d) \approx 0.56$ here as mentioned earlier. Combining $\tilde{t}_{\textrm{TP}} = Q \tilde{\tau} $ with the effective forces in the TP and PP stages together gives the total translocation time as $\tilde{\tau} =\tilde{L}_{\textrm{r}}^{\nu} \int_{0}^{N_{01}} \!\!\! \textrm{d} N_1 [ \tilde{R}_1 (N_1) + \tilde{\eta}_{\textrm{p}} ] / (G \tilde{F}_{\textrm{SP}})$. Using $\tilde{R}_1 (N_1) = A_{\nu}  N_1^{\nu} $, where $\nu = 3/4$ in 2D and $A_{\nu} = 1.1$ (from LD data), the scaling of the translocation time is
\begin{equation}
\tilde{\tau} = \frac{\tilde{L}_{\textrm{r}}^{\nu}}{G \tilde{F}_{\textrm{SP}}} 
\big[ \frac{ A_{\nu} N_{01}^{1+\nu} }{1+\nu} + N_{01} \tilde{\eta}_{\textrm{p}} \big],
\label{tau}
\end{equation}
where $G = -0.5 (c-a)^2 / (b+d) + c - d/2 \approx 11.25$. Equation~(\ref{tau}) reveals that the SP force and rod length exponents are $\beta = -1$ and $\gamma = \nu$, respectively and the translocation exponent varies between $1 < \alpha \leq 1+\nu$. The force and translocation exponents are in agreement with purely pore-driven translocation in the SS regime \cite{jalalJCP2014}. In Fig.~\ref{waiting}(b) we plot the translocation time as a function of the SP force (open orange squares from the top red horizontal axis) and the rod length (open turquoise circles from the bottom blue horizontal axis). Dashed red and dashed-dotted blue lines are guides to the eye. The SP force and rod length exponents obtained from IFTP theory are in good agreement with LD data.

%%%%%%%%%%%%%%%%%%%%%%%%%%%%%%%%%%%%%%%%%%%%%%%%%%%%%%%%%%%%%%%%%%%%%%%%%%%%%%%%%%%%%%%%%%%%%%%%%%%%%%%%%%%%%%%%%%%%%%%%%%%%%%%%%%%%%%%%%%%%%%%%%%%%%%%%%%%%%%%%%%%%%%%%%%%%%%%%%%%%%%%%%%%%%%%%%%%%%%%%%%%%%%%%%%%%%%%%%%%%%%%%%%%%%%%%%%%%%%%%%%%%%%%%%%%%%%%%%%%%%%%%%%%%%%%%%%%%%%%%%%%%%%%%%%%%%%%%%%%%%%%%%%%%%%%%%%%%%%%%%%%%%%%%%%%%%%%%%%%%%%%%%%%%%%%%%%%%%%%%%%%%%%%%%%%%%%%%%%%%%%%%%%%%%%%%%%%%%%%%%%%%%%%%%%%%%%%%%%%%%%%%%%%%%%%%%%%%

In summary, we have shown here that active rodlike particles in the {\it trans} side of a membrane can efficiently overcome entropic losses and facilitate translocation of a polymer chain through a nanopore. The SP force induces a crowding effect of the rods close to the membrane and the polymer, and as a net result there is an effective driving force making translocation possible even without explicit driving. We have used a combination of LD simulations and IFTP theory in the SS regime to characterise the waiting time distribution $w$ and the average translocation time $\tau$. Neglecting the explicit contribution of $N_{02}$ to the dynamics allows us to derive a scaling form for $\tilde \tau$ as a function of $N_{01}$, SP force and rod length. The scaling exponents for $\tau$ and the SP force are in agreement with those of the purely pore-driven translocation case. Our work gives new insight into our knowledge about the role of APs in living cells that may assist translocation of biomolecules, and may be used as a method to control translocation dynamics which is crucial for DNA sequencing applications.

%%%%%%%%%%%%%%%%%%%%%%%%%%%%%%%%%%%%%%%%%%%%%%%%%%%%%%%%%%%%%%%%%%%%%%%%%%%%%%%%%%%%%%%%%%%%%%%%%%%%%%%%%%%%%%%%%%%%%%%%%%%%%%%%%%%%%%%%%%%%%%%%%%%%%%%%%%%%%%%%%%%%%%%%%%%%%%%%%%%%%%%%%%%%%%%%%%%%%%%%%%%%%%%%%%%%%%%%%%%%%%%%%%%%%%%%%%%%%%%%%%%%%%%%%%%%%%%%%%%%%%%%%%%%%%%%%%%%%%%%%%%%%%%%%%%%%%%%%%%%%%%%%%%%%%%%%%%%%%%%%%%%%%%%%%%%%%%%%%%%%%%%%%%%%%%%%%%%%%%%%%%%%%%%%%%%%%%%%%%%%%%%%%%%%%%%%%%%%%%%%%%%%%%%%%%%%%%%%%%%%%%%%%%%%%%%%%%%

\begin{acknowledgments}
Computational resources from CSC - Center for Scientific Computing Ltd. are gratefully acknowledged.
T.A-N. has been supported in part by the Academy of Finland through its 
PolyDyna (no. 307806) and QFT Center of Excellence Program grants (no. 312298).
\end{acknowledgments}

%%%%%%%%%%%%%%%%%%%%%%%%%%%%%%%%%%%%%%%%%%%%%%%%%%%%%%%%%%%%%%%%%%%%%%%%%%%%%%%%%%%%%%%%%%%%%%%%%%%%
%%%%%%%%%%%%%%%%%%%%%%%%%%%%%%%%%%%%%%%%%%%%%%%%%%%%%%%%%%%%%%%%%%%%%%%%%%%%%%%%%%%%%%%%%%%%%%%%%%%%


\begin{thebibliography}{99}
\bibitem{dimensionless} The dimensionless quantities are denoted as $\tilde{Q} \equiv Q / Q_u$, where the denominator denotes the units of length, time, velocity, monomer flux, friction and force as $s_u \equiv \sigma$, $t_u \equiv \eta \sigma^2 / (k_{\rm B} T)$, $v_u \equiv \sigma/t_u = k_{\rm B} T/(\eta \sigma)$, $\phi_u \equiv k_{\rm B} T/(\eta \sigma^2)$, $\Gamma_u \equiv \eta$ and $f_u \equiv k_{\rm B} T/\sigma$, respectively.
\end{thebibliography}
\end{document}